\documentclass[preprint]{ptephy_v2}

\preprintnumber{2508.01449} 
\usepackage{hyperref}

\usepackage{amsmath} 
\usepackage{amsthm} 
\usepackage{hyperref} 
\usepackage{graphics} 
\usepackage{algorithmic} 
\usepackage{subfig} 
\usepackage{url} 
\begin{document}

\title{Boosting Sensitivity to $HH\to b\bar{b} \gamma\gamma$ with Graph Neural Networks and XGBoost}


\author[1,*]{Mohamed Belfkir}
\author[1,$\dagger$]{Mohamed Amin Loualidi}
\author[1,$\ddagger$]{Salah Nasri}
\affil[1]{Department of physics, United Arab Emirates University, Al-Ain, UAE}
\affil[*]{\email{\href{mailto:m_belfkir@uaeu.ac.ae}{m\_belfkir@uaeu.ac.ae}}}
\affil[$\dagger$]{\email{\href{mailto:ma.loualidi@uaeu.ac.ae}{ma.loualidi@uaeu.ac.ae}}}
\affil[$\ddagger$]{\email{\href{mailto:snasri@uaeu.ac.ae}{snasri@uaeu.ac.ae}, \href{mailto:salah.nasri@cern.ch}{salah.nasri@cern.ch} }}

\begin{abstract}%
In this paper, we explore the use of advanced machine learning (ML) techniques to enhance the sensitivity of double Higgs boson searches in the \( HH \to b\bar{b}\gamma\gamma \) decay channel at $\sqrt{s} = $ 13.6 TeV. Two ML models are implemented and compared: a tree-based classifier using XGBoost, and a geometrical-based graph neural network classifier (GNN). We show that the geometrical model outperform the traditional XGBoost classifier improving the expected 95\% CL upper limit on the double Higgs boson production cross-section by 28\%. Our results are compared to the latest ATLAS experiment results, showing significant improvement of both upper limit and Higgs boson self-coupling ($\kappa_{\lambda}$) constraints.
\end{abstract}

\maketitle

\section{Introduction}
\label{sec:intro}
Since the discovery of the Higgs boson in 2012 by the ATLAS and CMS experiments at the Large Hadron Collider (LHC) \cite{PhysRevLett.13.321, HIGGS1964132, PhysRevLett.13.508, Aad_2012, Chatrchyan_2012}, significant progress has been made in measuring its fundamental properties, including mass, spin, production cross-sections, and couplings to both bosons and fermions. These measurements have been achieved with high precision, confirming the consistency of the discovered Higgs boson with the predictions of the Standard Model (SM). However, the structure of the Higgs potential itself — particularly the nature of the Higgs boson self-coupling — remains largely unconstrained.
At low energies, the Higgs field potential can be parameterized, after electroweak symmetry breaking, as:
\begin{equation}
    \label{V}
    V(H) = \frac{1}{2}m_H^2 H^2 + \lambda_{3} v H^3 + \frac{\lambda_{4}}{4} H^4,
\end{equation}
where $m_H$ is the Higgs boson mass, precisely measured to be 125.11 $\pm$ 0.11 GeV \cite{Higgs_mass_ATLAS, Higgs_mass_CMS}, and $v \approx$ 246 GeV is the vacuum expectation value of the Higgs field. The parameters $\lambda_3$ and $\lambda_4$ represent the trilinear and quartic Higgs boson self-couplings, respectively. In the SM, the value of \( \lambda_{3} \approx 0.13 \) \cite{Li:2019jba}. The trilinear self-coupling \( \lambda_{3} \) governs the shape of the Higgs potential and plays a central role in understanding electroweak symmetry breaking and vacuum stability. It is also closely linked to early Universe cosmology and the nature of the electroweak phase transition ~\cite{PhysRevD.110.030001,PhysRevD.9.3357,Laine:471776}. While the quartic self-coupling $\lambda_4$ is beyond the direct reach of the LHC, the trilinear coupling can be experimentally probed through measurements of double Higgs boson production. However, current constraints on $\lambda_3$ remain subject to large uncertainties~\cite{arXiv:2407.14716, ATLAS:2022jtk, CMS:2024awa}, making it a key target for Beyond the SM (BSM) physics scenarios that predict deviations from the SM. These deviations are commonly parameterized using the dimensionless modifier $\kappa_{\lambda} = \lambda_3^\text{BSM} / \lambda_3^\text{SM}$. 

In this paper, we study the enhancement of the double Higgs boson search sensitivity using advanced ML algorithms. In particular, we focus on double Higgs boson production in the $HH \to b\bar{b}\gamma\gamma$ decay channel, which benefits from a clean final-state signature and good mass resolution, despite its relatively small branching ratio of approximately 0.2644\% for a Higgs boson mass of 125 GeV \cite{CERNYellowReportPageBR}.
Over the past decade, ML have played an increasingly important role in different physics analyses \cite{hepmllivingreview, Shanahan:2022ifi, Duarte:2024lsg, Choudhury:2024crp}, offering significant improvement in signal-to-background discrimination. To exploit these capabilities, we explore state-of-the-art ML techniques, including  traditional tree-based models \cite{Wu:2025jza} and GNNs \cite{Mullin:2019mmh, Guo:2020vvt, Sutcliffe:2025arn, Sahu:2024sts}, to improve the discrimination between signal and background processes. These algorithms are trained on a comprehensive set of high-level kinematic variables and event-level features, with GNNs additionally utilizing the geometric and topological information of reconstructed physics objects. 

By leveraging the geometric representation of double Higgs boson events, this paper demonstrates that GNN has an advantage in extracting complex spatial and topological correlations among final-state particles. Specifically, the trained GNN model achieves an improvement of nearly 28\% in sensitivity to double Higgs boson production compared to traditional tree-based classifiers. In addition, the GNN exhibits greater stability in the presence of limited background statistics and systematic uncertainties. 

The paper is organized as follows: in Sec.~\ref{sec:HH}, we introduce briefly the double Higgs boson production at the LHC. Sec.~\ref{sec:MC} describes the Monte Carlo event generation workflow. In Sec.~\ref{sec:event}, we summarize the physics object definitions and the event pre-selection strategy. The machine learning algorithms and training strategies used in this study are detailed in Sec.~\ref{sec:ML}. Our statistical analysis and results are presented in Sec.~\ref{sec:result}, and conclusions are given in Sec.~\ref{sec:conc}.
\section{Double Higgs production at the LHC}
\label{sec:HH}
\begin{figure}[ht]
    \centering
    \subfloat[a][]{\includegraphics[width=0.35\linewidth]{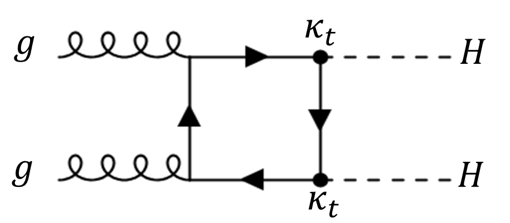}}
    \subfloat[b][]{\includegraphics[width=0.4\linewidth]{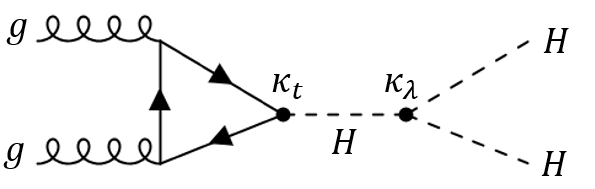}}\\
    \subfloat[c][]{\includegraphics[width=0.35\linewidth]{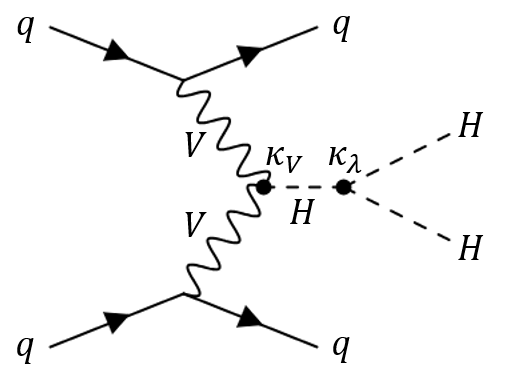}}
    \subfloat[d][]{\includegraphics[width=0.3\linewidth]{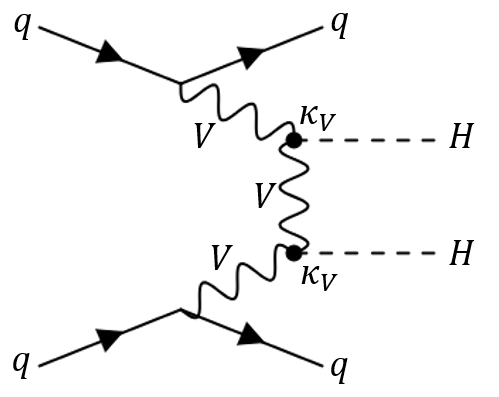}}
    \subfloat[e][]{\includegraphics[width=0.3\linewidth]{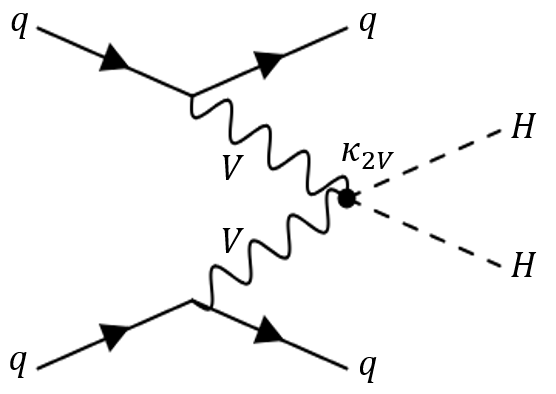}}
    \caption{Leading-order Feynman diagrams for the dominant ggF production (a and b), and VBF production (c-e) of double Higgs boson.}
    \label{fig:feyn}
\end{figure}

At the LHC, Higgs boson pair production is predominantly mediated by gluon-gluon fusion (ggF), proceeding via loop-level processes involving heavy quarks, primarily the top quark. This production mode is characterized by a destructive interference between the box diagram (Figure~\ref{fig:feyn}(a)) and the triangle diagram (Figure~\ref{fig:feyn}(b)), the latter of which is sensitive to the trilinear Higgs self-coupling making it highly sensitive to deviations in $\kappa_\lambda$. The sub-leading contribution arises from the vector boson fusion (VBF) mode, which is sensitive not only to the trilinear coupling $\kappa_\lambda$ (Figure \ref{fig:feyn} (c)), but also to the Higgs–vector boson (VVH) (Figure \ref{fig:feyn} (d)) and quartic Higgs-vector boson (VVHH) couplings (Figure \ref{fig:feyn} (e)). Deviations in the VBF topology are parameterized by the coupling modifiers $\kappa_V$ and $\kappa_{2V}$ as shown in the Feynman diagrams (d) and (e) of Figure \ref{fig:feyn}.

In the SM, the next-to-next-to-leading order (NNLO) cross-section for ggF Higgs boson pair production at $\sqrt{s} = 13.6~\text{TeV}$ is calculated to be 0.3413 pb with a theoretical uncertainty of $\pm 2.3\%$ from parton distribution functions (PDFs) and the strong coupling constant $\alpha_s$, and an asymmetric uncertainty of $ ^{+6\%}_{-23\%} $ due to renormalization/factorization scale and top quark mass variations~\cite{Dreyer_2020}. The subdominant VBF mode has a much smaller cross-section, computed at next-to-next-to-next-to-leading order (N3LO) as 0.001874 pb with an uncertainty of $\pm 2.7\% $ from PDFs+$\alpha_s$ and a negligible scale variation of $^{+0.05\%}_{-0.03\%} $~\cite{Dreyer_2020}. Despite the small cross-sections in the SM, BSM scenarios predict sizable enhancements in the di-Higgs production rate, particularly through modifications to the Higgs self-coupling and Higgs-vector boson vertices. This is because extended Higgs sectors in BSM naturally predict additional scalar states, including heavy Higgs bosons that can decay into final states containing more than one Higgs boson. Therefore, Higgs pair production is regarded as a promising channel for exploring new BSM physics. Prominent examples for such models include two-Higgs-doublet models (2HDMs) \cite{Lee:1973iz, Branco:2011iw, Haber:2015pua, Kling:2016opi}, their extensions with a scalar singlet (2HDM+S) \cite{Chen:2013jvg, Muhlleitner:2016mzt, Chalons:2012qe} and the two-real-singlet model (TRSM) \cite{Robens:2019kga}, each offering a rich phenomenology of scalar resonances. Supersymmetric frameworks inherently require such extensions: the minimal supersymmetric Standard Model (MSSM) \cite{Djouadi:2005gj, Gunion:1984yn, Gunion:1986nh, Degrassi:2002fi} incorporates a 2HDM-type Higgs sector, while the next-to-minimal supersymmetric extension (NMSSM) \cite{Maniatis:2009re, Ellwanger:2009dp, King:2012is} introduces an additional gauge-singlet scalar, effectively realizing a 2HDM+S structure. This would result in an additional source of Higgs boson pair production via resonant topologies, which are not predicted by the SM. In this paper, however, we restrict our analysis to the non-resonant production mode, where deviations in the Higgs self-coupling parameter manifest through modifications to the production cross-section. The variation of the ggF and VBF cross-section as a function of $\kappa_\lambda$ is typically parameterized using quadratic parameterizations. The cross-sections for these processes are given by \cite{Dreyer_2020}:
\begin{equation}
    \sigma_{pp \to HH}(\kappa_{\lambda}) = 75.7617 - 53.2855 \times \kappa_{\lambda} + 11.6126 \times \kappa_{\lambda}^{2} \ \text{[fb]},
\end{equation}
for the ggF channel, and
\begin{equation}
        \sigma_{pp \to HHjj}(\kappa_{\lambda}) = 0.0032 - 0.0029 \times \kappa_{\lambda} + 0.00093 \times \kappa_{\lambda}^{2} \ \text{[fb]},
\end{equation}
for the VBF channel. The VBF cross-section is generally parameterized including also $\kappa_{2V}$ and $\kappa_{V}$ \cite{bbyy_legacy}. However, in this study we focus on $\kappa_{\lambda}$ variations keeping all other couplings fixed to their SM values. 
\section{Monte Carlo event generation}
\label{sec:MC}
Our main goal is to investigate the potential enhancements offered by advance ML techniques in the search for double Higgs boson production in the $H(\to b\bar{b})H(\to\gamma\gamma)$ decay channel at the LHC. The analysis is based on simulation of proton-proton collision at a center-of-mass energy of 13.6 TeV, and an integrated luminosity of 168 fb$^{-1}$, corresponding to the amount of data collected by the ATLAS experiment during the partial Run-3 data-taking period form 2022 to 2024. The final state under consideration consists of two isolated photons and two resolved $b$-tagged jets. This channel is often referred to as "golden" due to the large branching ratio of $H\to b\bar{b}$, the excellent resolution of the di-photon invariant mass $m_{\gamma\gamma}$ distribution (invariant mass of two photons) and the well-understood continuum QCD background from $\gamma\gamma$ + jets processes. The main backgrounds mimicking the $b\bar{b} \gamma\gamma$ final state are categorized into two types:
\begin{itemize}
    \item \textbf{Resonant background}: This arises from single Higgs boson production followed by the decay $H\to\gamma\gamma$. Four production modes are considered: gluon-gluon fusion (ggF), vector boson fusion (VBF), associated production with a Z boson via both $qq \to ZH$ and $gg \to ZH$, and associated production with top quarks $t\bar{t}H$. 
    \item \textbf{Non-resonant background} (continuum background): This corresponds to direct QCD production of di-photon events accompanied by jets, which does not involve a Higgs resonance.
\end{itemize}
Although the continuum $\gamma\gamma$ + jets background dominates, it can be significantly suppressed by applying a selection requirement on the invariant mass $m_{\gamma\gamma}$ and additional kinematic constraints. 

The signal samples include both the dominant ggF and subdominant VBF production modes. The Monte Carlo simulation for the ggF HH process, with the Higgs self-coupling set to its SM value, is generated at next-to-leading order (NLO) using \texttt{Powheg-Box v2} \cite{PaoloNason_2004,Alioli_2010} with full NLO corrections including finite top quark mass effects \cite{Heinrich_2017, Heinrich_2019}. The generated events are further scaled using NNLO cross-section corrections with full top mass dependence at $m_H$ =  125.09 GeV, as provided by the LHC Higgs Working Group 4 (WG4) \cite{Grazzini_2018, Baglio_2021}.  The SM VBF HH signal samples are generated at LO using \texttt{MadGraph5\_aMC\@NLO v3.3.0} \cite{Alwall_2014}. VBF HH events are normalized using theoretical cross-sections evaluated at $m_{H} = $ 125.09 GeV, at N3LO QCD and NLO electroweak (EW) corrections \cite{Dreyer_2018,Dreyer_2020}.  \\
The single Higgs boson background events are generated using \texttt{Powheg-Box v2}, following the same methodology employed for the ggF double Higgs signal simulation. The continuum di-photon background, which arises primarily from non-resonant QCD-induced $\gamma\gamma$ + jets processes, is simulated at LO using \texttt{MadGraph5\_aMC@NLO v3.3.0}. The generation includes matrix elements with up to one and two additional partons in the final state, enabling a more accurate modeling of the jet multiplicity spectrum. Table \ref{tab:mc_table} summarizes the simulated processes with respective number of events and cross-sections.
\begin{table}[hbt!]
    \centering
    \begin{tabular}{lccc}
    \hline\hline
     Physics process &  MC Generator & Number of entries & Cross-section [pb] \\
     \hline
    $pp \to HH$ (SM) & Powheg & 50k & 0.3413 \cite{Dreyer_2020}\\
    $pp \to HHjj$ (SM) & MadGraph & 100k & 0.001874 \cite{Dreyer_2020}\\
    \hline 
    $pp \to H$ & Powheg & 500k & 52.17 \cite{CERNYellowReportPageBR}\\
    $pp \to Hjj$ & Powheg & 550k & 4.075 \cite{CERNYellowReportPageBR}\\
    $qq \to ZH $ & Powheg & 50k & 1.834 $\times 10^{-3}$ \cite{CERNYellowReportPageBR}\\
    $gg \to ZH $ & Powheg & 100k & 3.087 $\times 10^{-4}$ \cite{CERNYellowReportPageBR}\\
    $pp \to t\bar{t}H$ & Powheg & 100k & 5.688 $\times 10^{-1}$ \cite{CERNYellowReportPageBR}\\
    \hline 
    $\gamma\gamma $ + jets & MadGraph & 2.5M & 48.1 \\
    \hline\hline
    \end{tabular}
    \caption{Summary of generated signal and background processes.}
    \label{tab:mc_table}
\end{table}

Both generated signal and background events are processed through \texttt{Pythia 8.186}~\cite{pythia8} for the simulation of the parton shower evolution, hadronization, and modeling of the full decay chain, including the description of the underlying event. To emulate the detector response, the events are passed through the \texttt{Delphes} fast simulation framework~\cite{delphes}, using a dedicated ATLAS detector configuration card. The card has been updated to reflect the most recent performance calibrations, including realistic parameterizations of tracking efficiencies, calorimeter resolutions, jet energy scale and resolution corrections, and $b$-tagging efficiency and mis-tag rates.
\section{Object definitions and event reconstruction}
\label{sec:event}
\subsection{Object definitions}
The characteristic event topology for the $HH \to b\bar{b} \gamma\gamma$ signal consists of two isolated photons and two $b$-tagged jets in the final state. To ensure accurate and consistent object reconstruction and selection, this section provides a detailed description of the physics object definitions employed in the analysis, including photons, jets, $b$-tagging criteria, and event-level selection.

\begin{itemize}
    \item \textbf{Photons}: reconstructed from energy deposits in the electromagnetic calorimeter using tower-based clustering algorithms as implemented in the \texttt{Delphes} fast simulation framework. Reconstructed photon candidates are required to have a transverse momentum $ p_T > 20~\text{GeV}$ and lie within the pseudorapidity range $|\eta| < 2.37$, excluding the ATLAS calorimeter barrel-endcap transition region defined by $ 1.37 < |\eta| < 1.52 $, where the detector performance is known to degrade due to reduced granularity and inactive material~\cite{ATLAS:2022hmt}. To avoid duplicate photon reconstruction and ensure the selection of spatially distinct candidates, a self-overlap removal requirement is applied. In cases where two photon candidates are separated by less than $\Delta R < 0.01$, only the leading (highest $p_T $) photon is retained. Moreover, photon candidates are required to pass the Tight identification working point (WP) emulated assuming the published photon identification efficiencies measured using Run-3 data collected in 2022-2024 at 13.6 TeV available in Ref. \cite{photonID}.

    \item \textbf{Jets}: reconstructed from energy deposits in the calorimeter using the anti-$k_t$ algorithm~\cite{Cacciari_2008}, with a radius parameter $R = 0.4$, implemented via the \texttt{FastJet} package~\cite{Cacciari:2011ma}. The inputs to the clustering are calorimeter towers, which approximate the energy flow in the detector. Reconstructed jets are required to satisfy a transverse momentum threshold of $p_T > 25~\text{GeV}$ and a rapidity requirement of $|y| < 4.5$, ensuring that jets are within the acceptance of the ATLAS calorimeter and are well-reconstructed. Jets originating from $b$-quarks ($b$-jets) are identified by emulating the performance of the ATLAS $b$-tagging algorithm at the 85\% efficiency WP, consistent with the latest ATLAS Run-3 recommendations \cite{btagGN2}. This corresponds to a mis-identification rate of 0.17 for \( c \)-jets and 0.01 for light-flavor jets, thereby allowing the background to include contributions from mis-tagged jets. The $b$-tagged jets are required to be within the tracking acceptance ($|\eta| < 2.5)$.
    \item \textbf{Leptons (electrons or muons)}: reconstructed using the particle-flow tracks collection. Muons are required to have a transverse momentum $p_T$ greater than 10 GeV and the pseudorapidity within the range of $|\eta| < $  2.7. Conversely, electrons are required to have $p_T > 10 \ \text{GeV}$ and $|\eta|< 2.47$ (excluding electrons failing in the calorimeter transition region).
\end{itemize}
\subsection{Event selections}
\label{selection}
Events are selected using a di-photon trigger designed to capture events containing two energetic photons. The di-photon trigger requires the leading and sub-leading photons to have transverse energies exceeding 35 GeV and 25 GeV, respectively. The trigger efficiency is implemented using publicly available efficiency measurements derived from 2022 data at 13.6 TeV, as documented in Ref.~\cite{trig}. An object cleaning procedure is applied to suppress overlaps between reconstructed objects and to reject non-isolated candidates.

Following Refs.~\cite{bbyy_legacy,atlas_results, bbyy_run2}, the two leading photons are required to have an invariant mass in the range $105~\text{GeV} < m_{\gamma\gamma} < 160~\text{GeV}$. Additionally, their transverse momenta must satisfy $p_T^{\gamma_1} > 0.35 \, m_{\gamma\gamma}$ and $p_T^{\gamma_2} > 0.25 \, m_{\gamma\gamma}$, respectively. This cuts scales with $m_{\gamma\gamma}$ so that the selection efficiency is flat across the mass spectrum. This avoids biasing the background shape near the Higgs peak \cite{ATLAS:2014euz}. The two photons are used to reconstruct the $H\to\gamma\gamma$ candidates. For the $H \to b\bar{b}$ reconstruction, events are required to contain at least two \( b \)-tagged jets, with the two leading $b$-jets taken as the $H \to b\bar{b}$ candidate.

To suppress background contributions from the semi-leptonic decay of the $t\bar{t}H$ process, events containing leptons (electrons or muons) are vetoed. Moreover, to reduce the hadronic \( t\bar{t}H \) contamination, events are required to contain no more than six central jets.
In addition to the baseline event selection, two VBF-tagged jets are identified—when present—as the two highest-$p_T$ jets excluding the $b$-tagged jets used in the reconstruction of the $H \to b\bar{b}$ candidate. These jets are intended to capture the characteristic topology of vector boson fusion, typically associated with forward-scattered quarks. However, the presence of VBF jets is not a strict requirement, and events lacking such jets are retained. It should be noted that this study does not define a dedicated VBF event category. The inclusion of such a category could significantly enhance the sensitivity to the $\kappa_{2V}$ coupling, and will be explored in a future analysis.

\begin{figure}
    \centering
    \subfloat[$m_{\gamma\gamma}$][$m_{\gamma\gamma}$]{\includegraphics[width=0.5\linewidth]{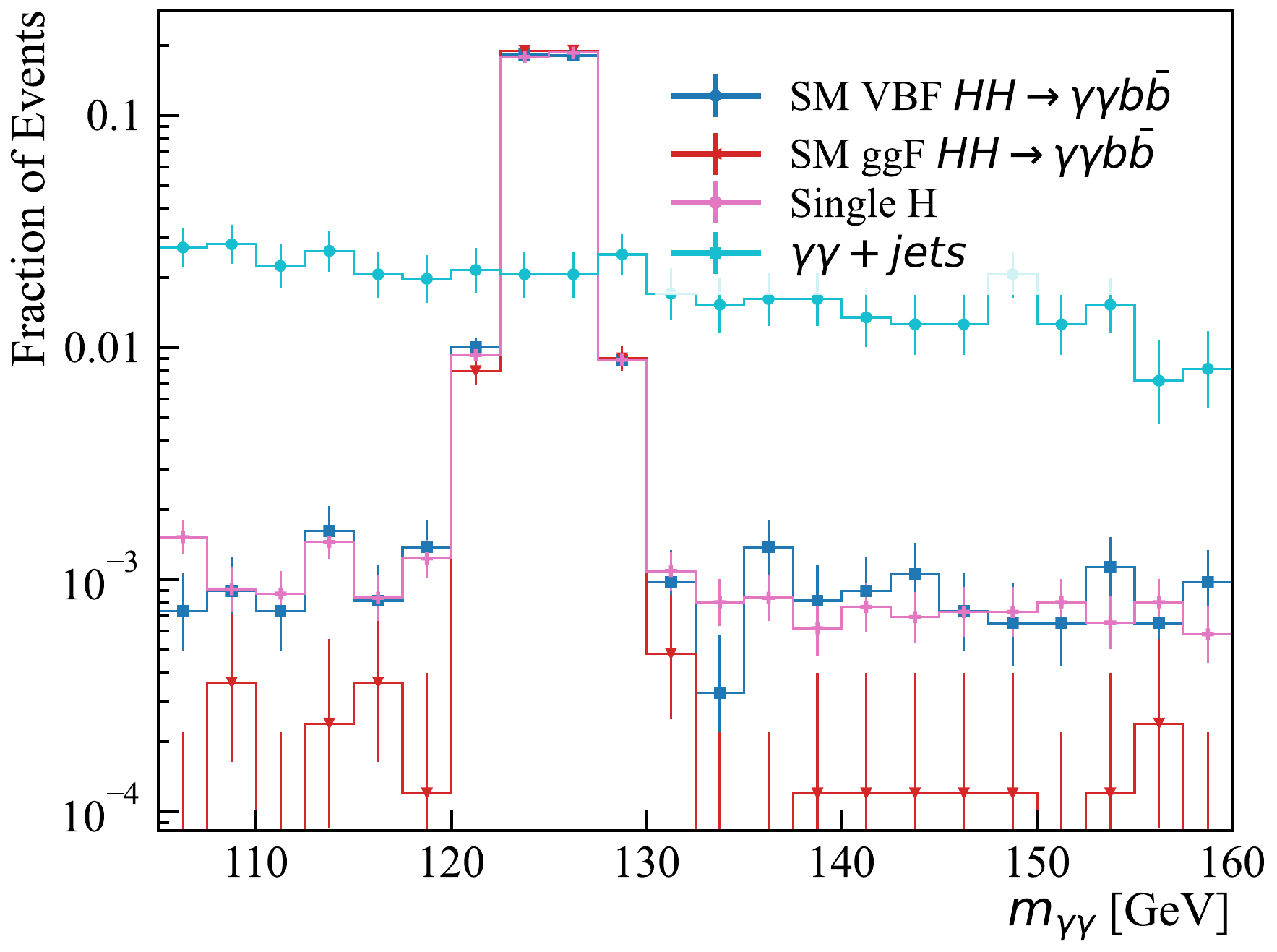}}
    \subfloat[$m_{bb}$][$m_{bb}$]{\includegraphics[width=0.5\linewidth]{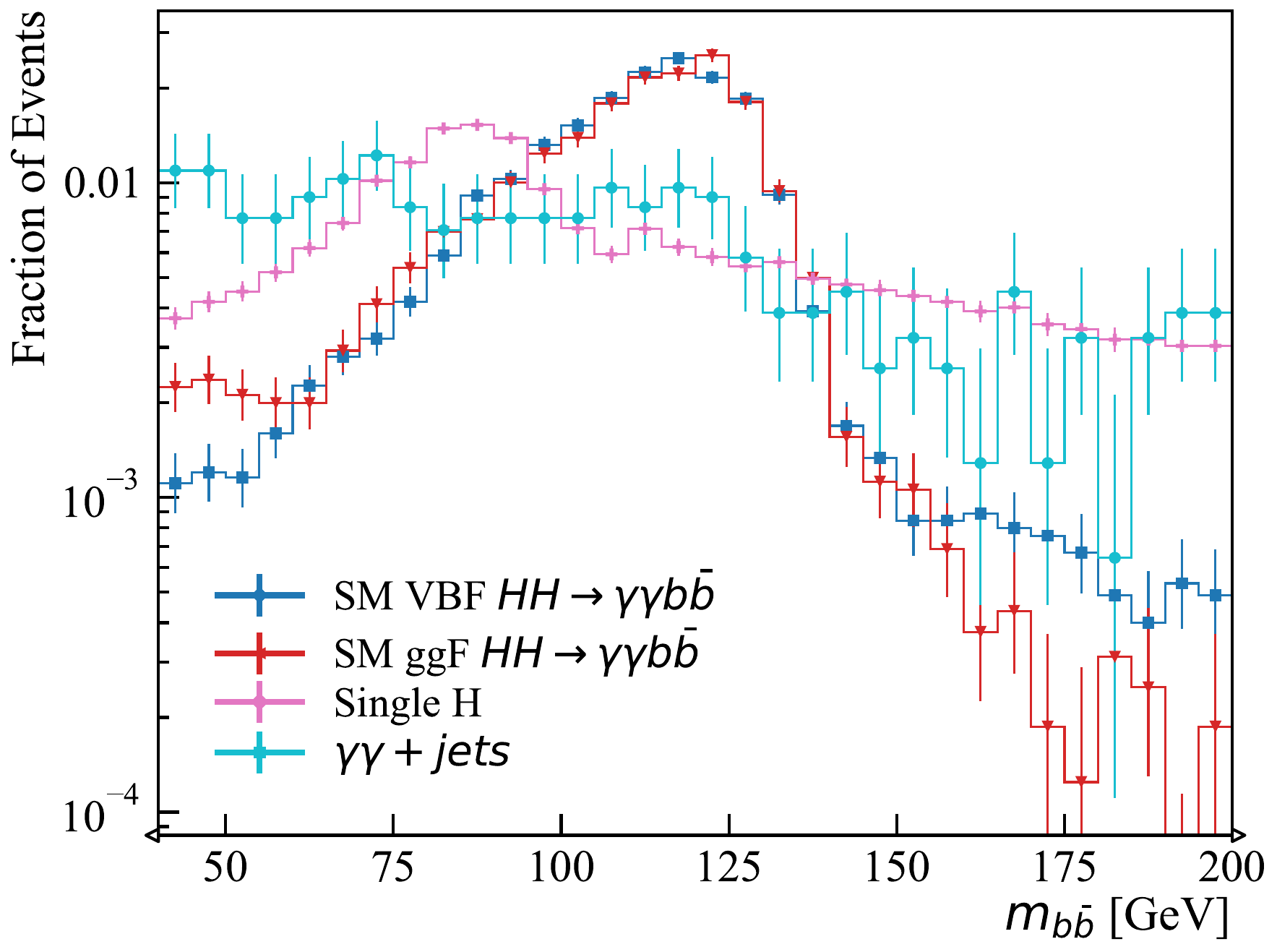}}
    \caption{Normalized distributions of (a) $m_{\gamma\gamma}$ and (b) $m_{bb}$ of selected events for SM signal, single Higgs background and the continuum $\gamma\gamma$ + jets. Distributions are normalized to unity.}
    \label{fig:myy_mbb}
\end{figure}

Figure~\ref{fig:myy_mbb} shows the invariant mass distributions of the di-photon and di-jets ($m_{bb}$) systems, as reconstructed for the SM signal processes—both ggF and VBF—as well as for single Higgs backgrounds and the continuum $\gamma\gamma +$ jets background. Due to the limited statistics available for the continuum diphoton background, the final results of the analysis—including signal significance and exclusion limits—are presented both with and without this background component to illustrate its potential impact. The relatively poor resolution of the $m_{bb}$ distribution can be attributed to several factors, including undetected leptons and neutrinos from semi-leptonic decays of $b$-hadrons, as well as out-of-cone effects arising from the large mass of the $b$-quark, which can cause part of the jet energy to fall outside the jet reconstruction cone. A dedicated $b$-jet energy calibration procedure, that goes beyond the scope of this paper, can improves the $m_{bb}$ resolution by approximately 20\%~\cite{bbyy_run2}.
\section{ML event categorization}
\label{sec:ML}
\subsection{Event pre-processing}
This section describes the event categorization strategy designed to enhance the sensitivity of the analysis to double Higgs boson production. As outlined previously, advanced machine learning techniques are employed to improve the separation between signal and background events beyond what is achievable with conventional cut-based selections. ML approaches have become increasingly prominent in different physics analyses, offering significant gains in signal-to-background discrimination through the exploitation of complex correlations between kinematic variables.

In this analysis, two distinct machine learning algorithms are explored for event classification. Prior to machine learning model training, a dedicated pre-processing pipeline is implemented to standardize the inputs and reduce detector-induced asymmetries. In particular, events passing the selection described in Section \ref{selection} are geometrically rotated in the transverse plane such that the leading photon is aligned with the beam axis. This transformation exploits the cylindrical symmetry of the ATLAS detector in \(\phi\), simplifying the learning task by removing arbitrary rotational variations and encouraging the network to focus on intrinsic event topology~\cite{Chatham_Strong_2020}. All other final-state objects, including the sub-leading photon and the two $b$-jets, are rotated accordingly to preserve their relative angular configuration. The event-rotation shows that the fitted signal strength is significantly improved by 12\%, and the $\kappa_{\lambda}$ 95\% confidence interval is reduced by roughly 10\%.
In addition, for neural network-based algorithms, all input variables are standardized such that their distributions are centered and follow a unit normal distribution. Specifically, each variable is transformed to have zero mean and unit variance. This standardization step improves numerical stability and accelerates the convergence of gradient-based optimization during training by ensuring that all features contribute on comparable scales~\cite{standarz}. The selected events are randomly divided into two statistically independent subsets: 75\% of the events are used for training the classifiers, while the remaining 25\% are reserved for evaluating model performance. To prevent training-induced biases and ensure an unbiased assessment of the classifier's ability to generalize, only the test subset is used to derive final analysis results. This data partitioning strategy ensures that the model performance is measured on events that were not seen during training, providing a realistic estimate of its performance. During classifier training, each event is assigned a weight corresponding to its generator-level cross section. This ensures that the relative contributions of different processes reflect their expected yields in data. To further account for the class imbalance between signal and background samples, an additional weighting is applied using class weights computed with the \texttt{compute\_sample\_weight} function from the \texttt{Scikit-Learn} package \cite{scikitlearn}. The total weight assigned to each event is given by the product of the generator-level event weight and the class weight. This approach helps mitigate training bias caused by the over representation of background events and ensures that the classifier is sensitive to both classes during optimization. 
\subsection{Tree-based classifier}
The first algorithm used in this analysis is a tree-based classifier, a class of machine learning models renowned for their robustness, interpretability, and state-of-the-art performance in signal-vs-background classification tasks \cite{Adhikary:2020fqf}. Tree-based methods, particularly boosted decision trees (BDTs), have become a standard tool in high-energy physics due to their ability to capture complex, non-linear correlations between variables while remaining relatively resistant to overfitting \cite{Hammad:2025ewr}. In our analysis, we implement the classifier using the \texttt{XGBoost} framework~\cite{xgboost}, which provides an efficient and scalable implementation of gradient-boosted trees. XGBoost has been widely adopted in experimental HEP analyses, offering strong performances even with limited hyperparameter tuning. The XGBoost is trained using 32 variables as listed in Table \ref{tab:xgboost_var}. As mentioned above, no standardization or normalization is applied to the input variables for the XGBoost training, since tree-based algorithms are insensitive to differences in variable scales.
\begin{table}[ht]
    \centering
    \begin{tabular}{ll}
    \hline\hline
        Objects & variables  \\
    \hline
       Photons  & $p_{T}$, $\eta$, $\phi$ \\
       $b$-jets & $p_{T}$, $\eta$, $\phi$ \\
       $\gamma\gamma$ system & $p_{T}$, $\eta$, $\phi$, $m$ \\
       $b\bar{b}$ system& $p_{T}$, $\eta$, $\phi$, $m$  \\
       $b\bar{b}\gamma\gamma$ system &  $p_{T}$, $\eta$, $\phi$, $m$ \\
       VBF-jets & $p_{T}$, $\eta$, $\phi$, $m_{jj}$, $\Delta \eta_{jj}$ \\
     \hline\hline  
    \end{tabular}
    \caption{XGBoost training features. Where $m_{jj}$ is the invariant mass of the two VBF jets and $\Delta \eta_{jj}$ is their $\eta$ separation.}
    \label{tab:xgboost_var}
\end{table}

\begin{figure}[hbt!]
    \centering
    \includegraphics[width=0.6\linewidth]{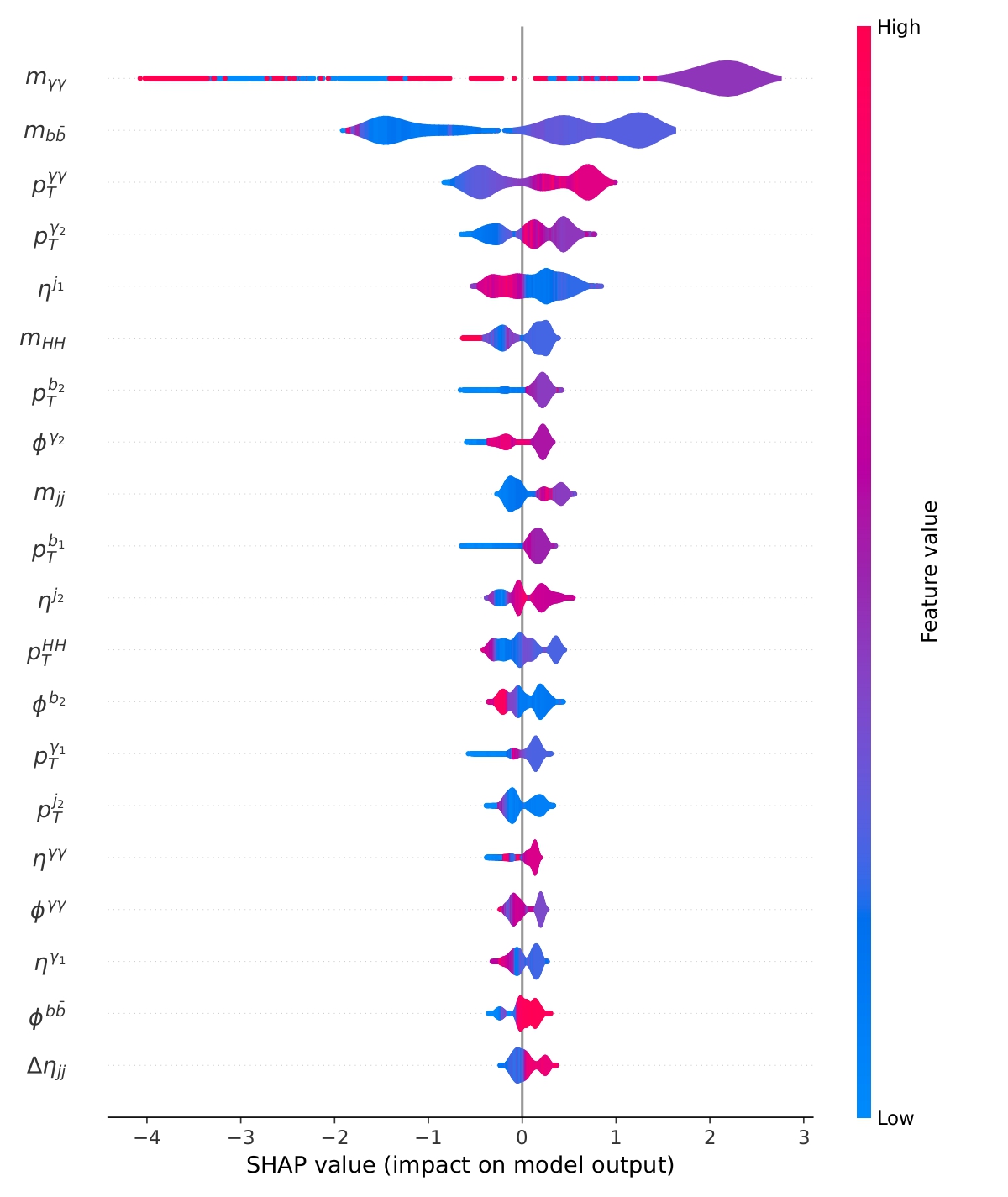}
    \caption{XGBoost feature importance ranking.}
    \label{fig:ranking}
\end{figure}

In Higgs physics analyses, the diphoton invariant mass \( m_{\gamma\gamma} \) is typically the most powerful discriminating variable for identifying Higgs boson decays to photon pairs. This holds true in the present analysis as well. In previous ATLAS studies such as~\cite{bbyy_legacy,bbyy_run2}, \( m_{\gamma\gamma} \) was excluded from the training features. This exclusion prevents the classifier from learning the discriminative power of \( m_{\gamma\gamma} \), which is instead reserved for the unbinned likelihood fit to extract the exclusion limit. In contrast, this analysis employs a binned likelihood fit to the output score of the classifier. This allows for a different treatment of input features. Since the final fit no longer relies on the invariant mass distribution, we choose to include \( m_{\gamma\gamma} \) in the training variable set. This enables the classifier to exploit its full discriminative power as demonstrated in Figure \ref{fig:ranking}.
The XGBoost algorithm includes a set of hyperparameters that significantly influence model performance and require careful tuning. In this analysis, we perform hyperparameter optimization using a random grid search approach, implemented via the \texttt{RandomizedSearchCV} method from the \texttt{Scikit-Learn} library \cite{scikitlearn}. Table~\ref{tab:xgboost_hyperparams} summarizes the hyperparameters \cite{xgboost-parameters} considered during the optimization process, along with the final values. The trained XGBoost model achieved an accuracy of 94.03\% on the 25\% test dataset.
\begin{table}[htb!]
    \centering
    \begin{tabular}{llcc}
    \hline\hline
    Parameter   &  Definition & Default value & Optimized value\\ 
    \hline 
    \texttt{min\_child\_weight} & Min. instance weight in a child node   & 1.0 & 7.0  \\
    \texttt{colsample\_bytree} & Fraction of features per tree    & 1.0 & 0.46 \\
    \texttt{scale\_pos\_weight} & Class weight balance factor    & 1.0 & 1.51 \\
    \texttt{max\_depth} & Maximum depth of the trees             & 6 &  5     \\
    \texttt{learning\_rate} & Step size shrinkage      & 0.3 & 0.049\\
    \texttt{subsample} & Fraction of training data per tree  & 1.0 & 0.93 \\
    \texttt{n\_estimators} & Number of boosting trees  & - & 919    \\
    \texttt{reg\_lambda} & L2 regularization term  & 1.0 & 3.17  \\
    \texttt{reg\_alpha} & L1 regularization term     & 0.0 & 0.41  \\
    \texttt{gamma}  & Min. loss reduction for a split & 0.0 & 0.17  \\
    \texttt{max\_delta\_step} & Max. step size in weight updates  & 0.0 &  6.87\\
    \hline\hline
    \end{tabular}
    \caption{Optimized XGBoost hyperparameters.}
    \label{tab:xgboost_hyperparams}
\end{table}
Figure \ref{fig:ranking} shows the relative importance of the input features in the XGBoost model and their contributions to its output, as quantified by SHAP (SHapley Additive exPlanations) values~\cite{SHAP}. 
Features are ranked by their overall importance, defined as the mean absolute SHAP value computed across all events. The invariant mass of the two photons is, as expected, at the top of the ranking. High values of $m_{\gamma\gamma}$ tend to shift the classifier output toward the signal class, reflecting the presence of a sharp mass peak near $125~\text{GeV}$ in signal events. Other highly ranked features include the invariant mass of the two $b$-jets $(m_{b\bar{b}})$, the transverse momentum of the di-Higgs system $(p_T^{HH})$, and the transverse momentum of the subleading photon. 
\subsection{Graph-based classifier}
While XGBoost has shown strong performance in signal-versus-background classification tasks using both low-level (e.g., particle momenta) and high-level (e.g., invariant mass, $\Delta \eta$) features, it processes each event as a flat vector of inputs. This vectorized approach overlooks the inherently structure of collider events, where complex topological relationships exist between particles, jets, and reconstructed objects. These patterns such as spatial correlations or decay chains are not explicitly modeled by tree-based algorithms. GNNs are specifically designed to operate on graph-structured data, making them well-suited for representing particle-level or event-level interactions \cite{Thais:2022iok, Sutcliffe:2025arn,Guo:2020vvt, Fuks:2025qgh} . By encoding particles as nodes and their interactions as edges, GNNs can learn directly from the event topology and improve classification by capturing both local and global context within each event. \\
In our analysis, each event is mapped to a bi-directional graph $\mathcal{G} = (\mathcal{V}, \mathcal{E})$, where nodes $v_{i} \in \mathcal{V}$ represent the reconstructed object including the two photons, two $b$-jets, two VBF-jets, as well as composed systems including $\gamma\gamma$, $b\bar{b}$, $\gamma\gamma b\bar{b}$. For each node, we include low-level kinematic features such as the $p_{T}$, $\eta$ and $\phi$. For composite systems, the invariant mass is also included as an additional node feature. Nodes are geometrically positioned in the $(\eta, \phi)$ space according to the spatial coordinates of the associated object. This spatial embedding allows the GNN to learn patterns related to detector geometry and angular separation between objects, an approach that has shown strong performance in collider-based event classification~\cite{mohammed2025geometricgnnschargedparticle, ExaTrkX:2020nyf, Anisha:2023xmh, Shlomi_2021,JetGNN}. The graph edges $e_{i} \in \mathcal{E}$, encode the physical and kinematic relationships among the reconstructed objects in the event. Symmetric edges are included between particles of the same type, such as:
\begin{equation}
    \gamma_1 \leftrightarrow \gamma_2, \quad  b_1 \leftrightarrow b_2, \quad VBF_1 \leftrightarrow VBF_2.
\end{equation}
Composite nodes such as the di-photon system $H_{\gamma\gamma}$, the di-$b$-jet system $H_{b\bar{b}}$, and the di-Higgs candidate $HH$ are connected hierarchically:
\begin{equation}
    \gamma_1 \leftrightarrow H_{\gamma\gamma} \leftrightarrow \gamma_2, \quad  b_1 \leftrightarrow H_{b\bar{b}} \leftrightarrow b_2, \quad H_{\gamma\gamma} \leftrightarrow HH \leftrightarrow H_{b\bar{b}}.
\end{equation}
An edge is included between $H_{\gamma\gamma}$ and $H_{b\bar{b}}$ to represent the interaction between the two Higgs candidates. All edges are duplicated in both directions to enable symmetric message passing across the graph. The $\Delta R$ between the nodes is included as edge attributes. Additionally, the graph includes a virtual node labeled \texttt{VirtVBF}, which connects to the two VBF jets as well as the $HH$ system, allowing the GNN to incorporate global VBF-related event features. This virtual node is designed to integrate global event-level VBF-specific information, such as the dijet invariant mass ($m_{jj}$) and the rapidity separation ($\Delta \eta_{jj} $) and it is placed at the origin in the $(\eta, \phi)$ coordinate space.

\begin{figure}[hbt!]
    \centering
    \subfloat[SM ggF $HH\to b\bar{b}\gamma\gamma$][SM ggF $HH\to b\bar{b}\gamma\gamma$]{\includegraphics[width=0.35\linewidth]{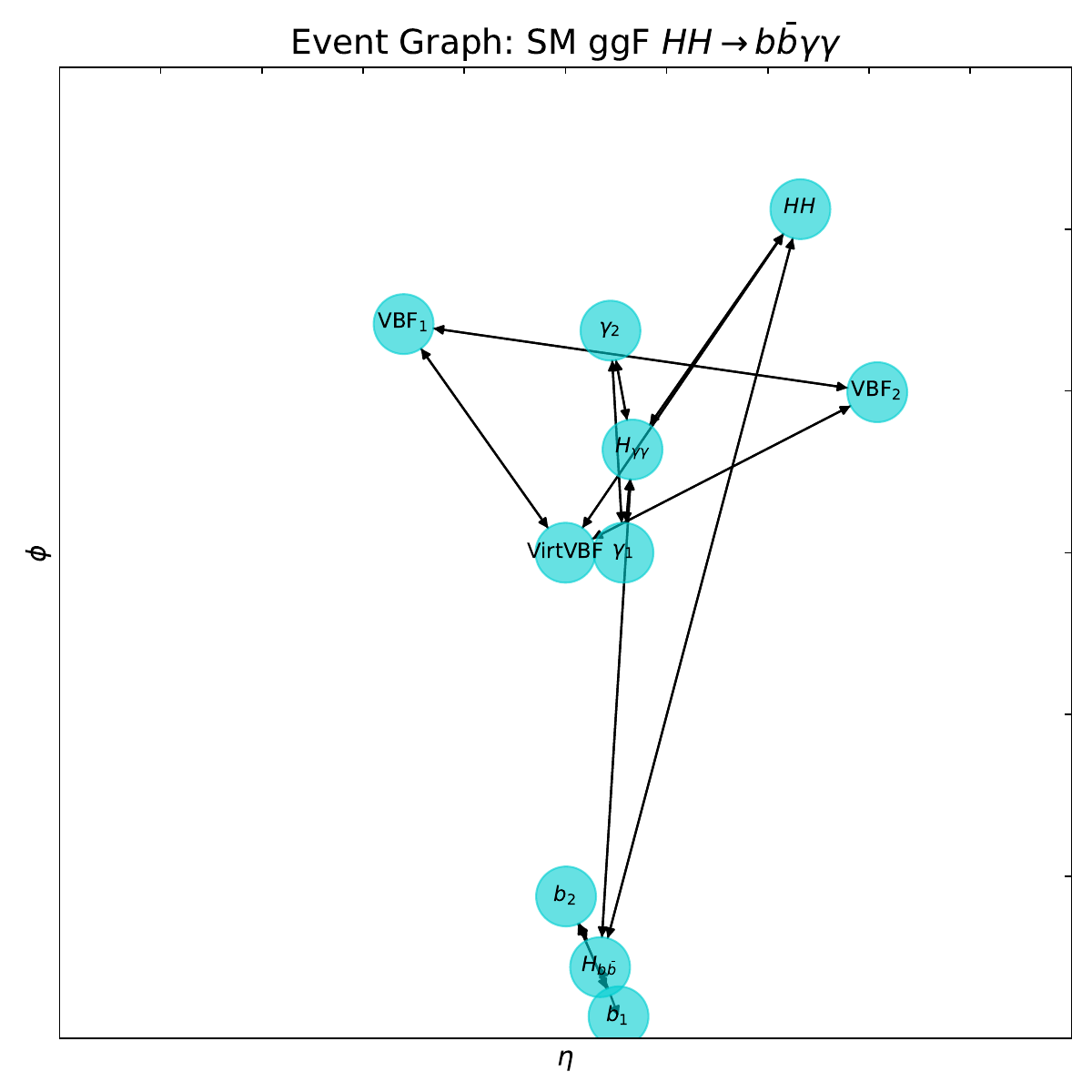}}
    \subfloat[SM VBF $HH\to b\bar{b}\gamma\gamma$][SM VBF $HH\to b\bar{b}\gamma\gamma$]{\includegraphics[width=0.35\linewidth]{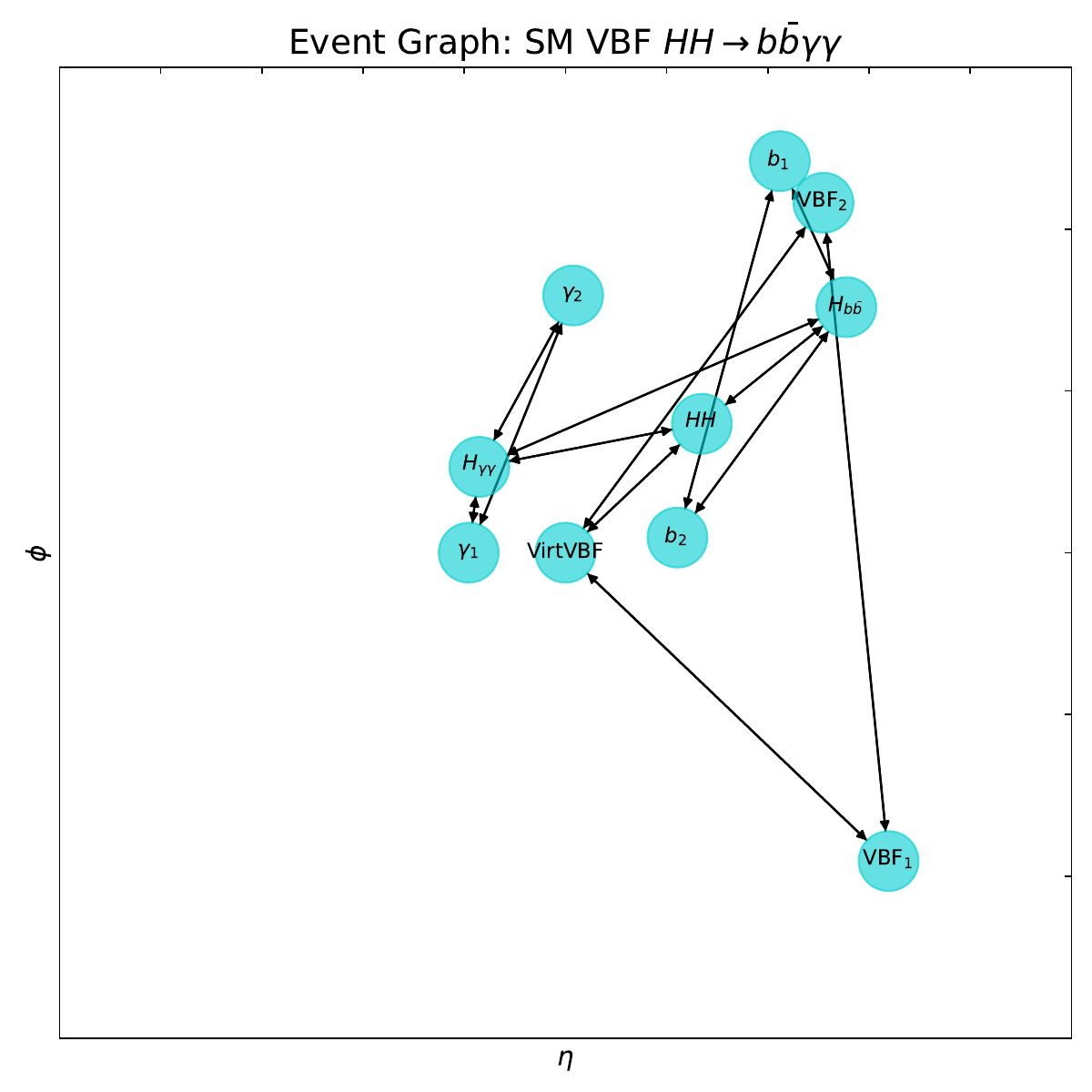}}\\
    \subfloat[$\gamma\gamma$ + jets][$\gamma\gamma$ + jets]{\includegraphics[width=0.35\linewidth]{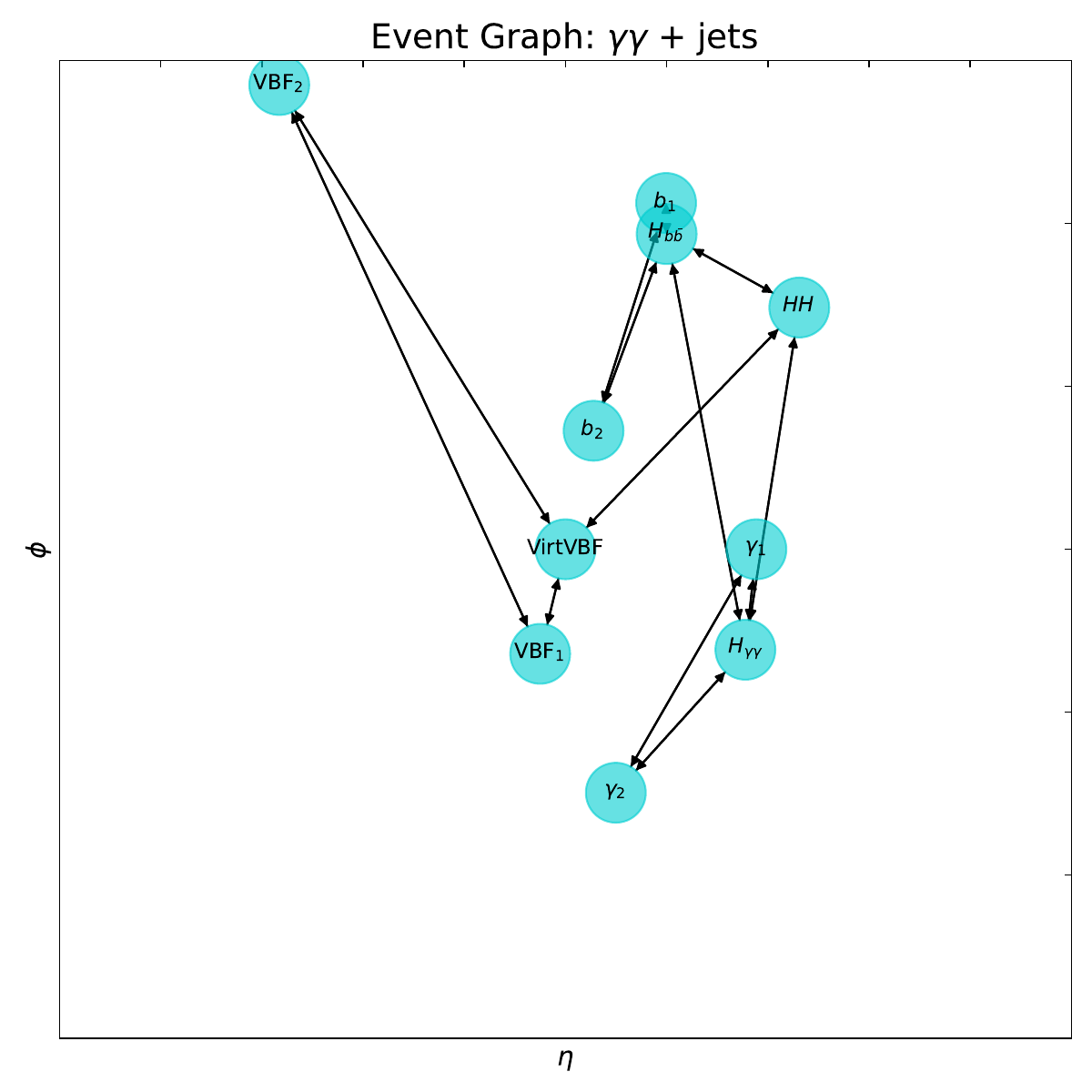}}
    \caption{Graphical representation of the event graph for (a) the SM ggF $HH\to b\bar{b}\gamma\gamma $, (b) the SM VBF $HH\to b\bar{b}\gamma\gamma$ and (c) the continuum $\gamma\gamma$ + jets.}
    \label{fig:graph}
\end{figure}

We developed the final graph structure through an iterative trial-and-error process, starting with fully connected graphs and gradually refining the topology based on performance evaluations. The resulting graph structure, which encodes only physically meaningful and hierarchically motivated connections, yielded the best performance. Figure~\ref{fig:graph} shows an illustration of the constructed event graphs for three processes: (a) ggF HH signal, (b) VBF HH signal, and (c) continuum background.
The created event graphs are processed using a message-passing graph neural network implemented with the \texttt{PyTorch Geometric} library \cite{pytorchgeo}. The model architecture consist of two edge-conditioned convolution (NNConv) layers, where the transformation of neighboring node features is modulated by a neural network applied to the edge attributes. In this study, the only edge-attribute used is the angular separation $\Delta R$ between nodes.

\begin{figure}[hbt!]
    \centering
    \includegraphics[width=0.8\linewidth]{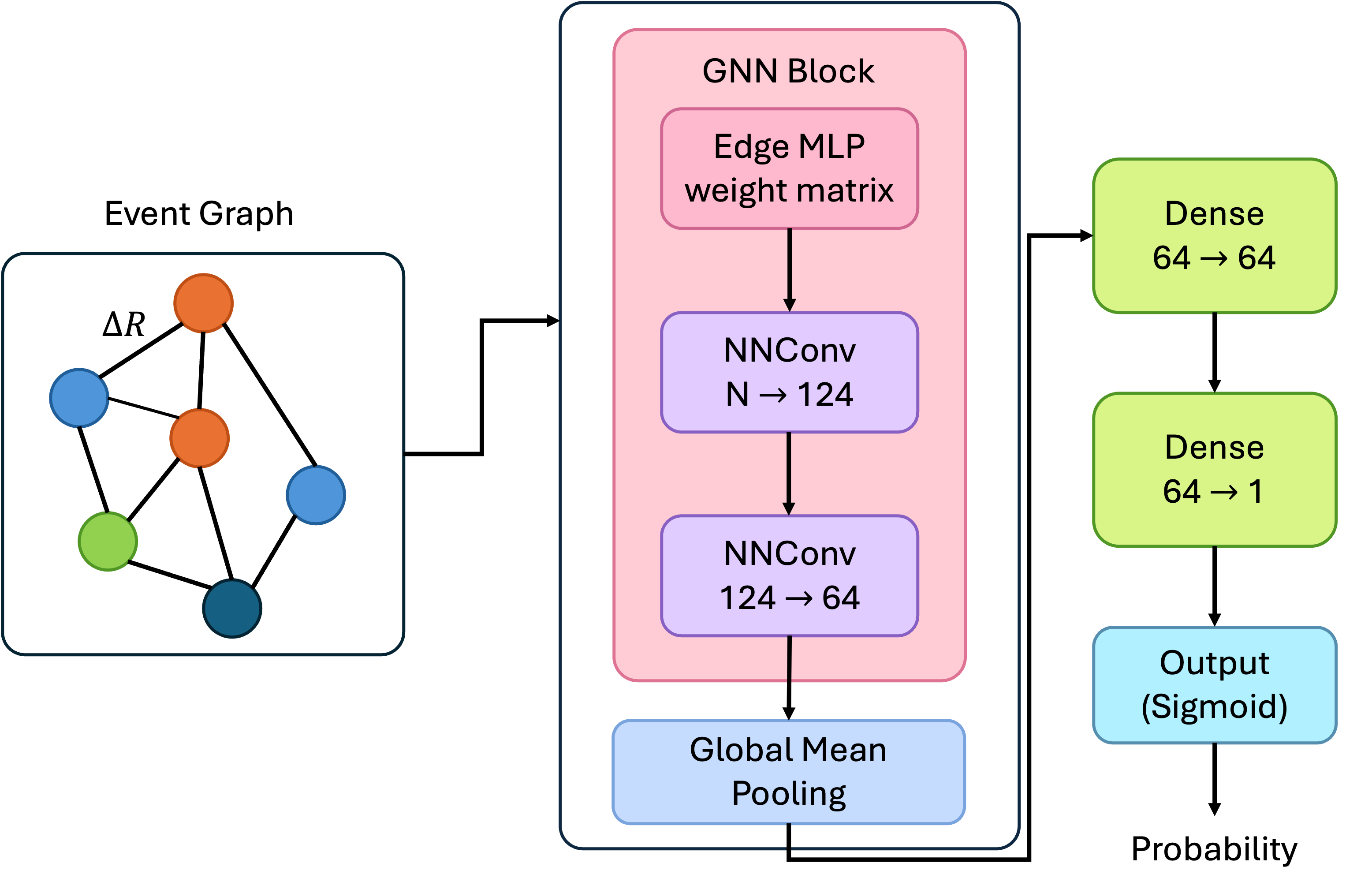}
    \caption{Overview of the geometry-aware GNN model.}
    \label{fig:GNN_schematic}
\end{figure}

The first edge network takes the $\Delta R$ and maps it through a two-multilayer perceptron (MLP) to a weight matrix of shape $N\times124$, allowing the first NNConv layer to project the $N$-dimensional input node features into a 124-dimensional latent space. The second edge network similarly maps $\Delta R$ to a $124\times64$ matrix, further reducing the dimensionality of the node embeddings in the second NNConv layer. Both convolutional layers use mean aggregation, which averages the transformed messages from neighboring nodes, providing stability and degree invariance. Following the convolutional layers, a global mean pooling operation aggregates the node-level features into a single event-level representation. This vector is then passed through two fully connected layers with 64 neurons and Rectified Linear Unit (ReLU) activation \cite{relu}. The final output is a scalar logit, which is converted into a probability of being a signal event via a sigmoid activation function. A schematic overview of the network is shown in Figure \ref{fig:GNN_schematic}.

The model is trained using the Adam optimizer with a learning rate of $10^{-3}$, and the binary cross entropy loss is used as the objective function \cite{elharrouss2025lossfunctionsdeeplearning}. Training is performed for up to 50 epochs, with early stopping based on validation loss: if no improvement is observed over 5 consecutive epochs, training is ended to prevent overfitting. A batch size of 1024 events is used for stable gradient estimates. The GNN model achieved an accuracy of 95.42\% on the test dataset.
\section{Results and discussion}
\label{sec:result}
\begin{figure}[hbt!]
    \centering
    \includegraphics[width=0.7\linewidth]{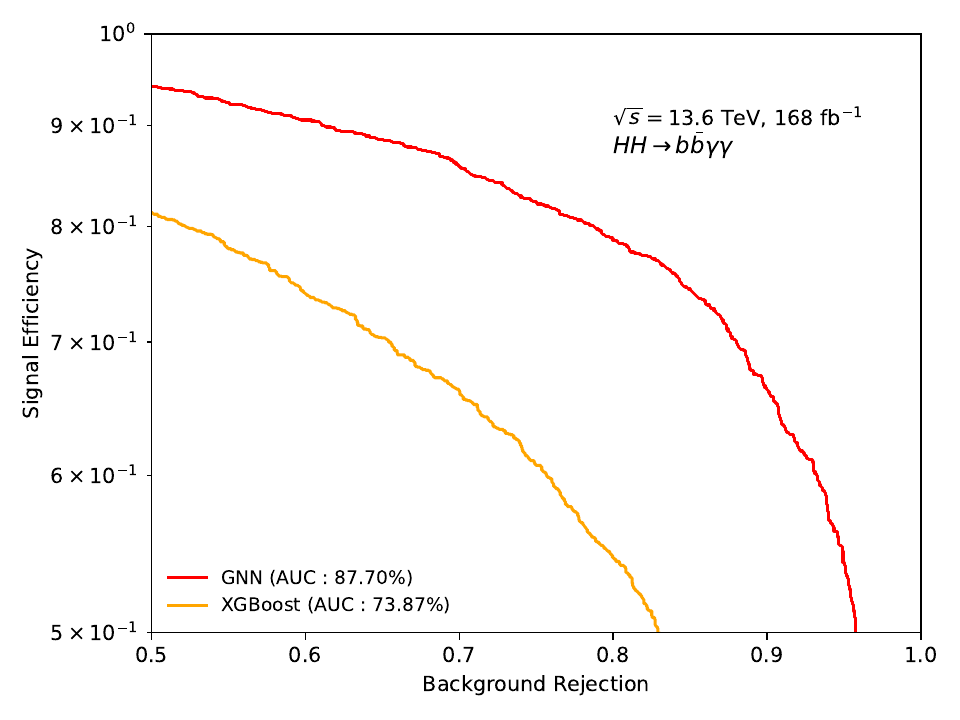}
    \caption{Weighted ROC curves for GNN model (red) and the XGBoost classifier (orange). The ROC curves are computed using event weights  on the test datasets.}
    \label{fig:ROC}
\end{figure}

\begin{figure}[hbt!]
    \centering
    \subfloat[GNN Classifier][GNN Classifier]{\includegraphics[width=0.5\linewidth]{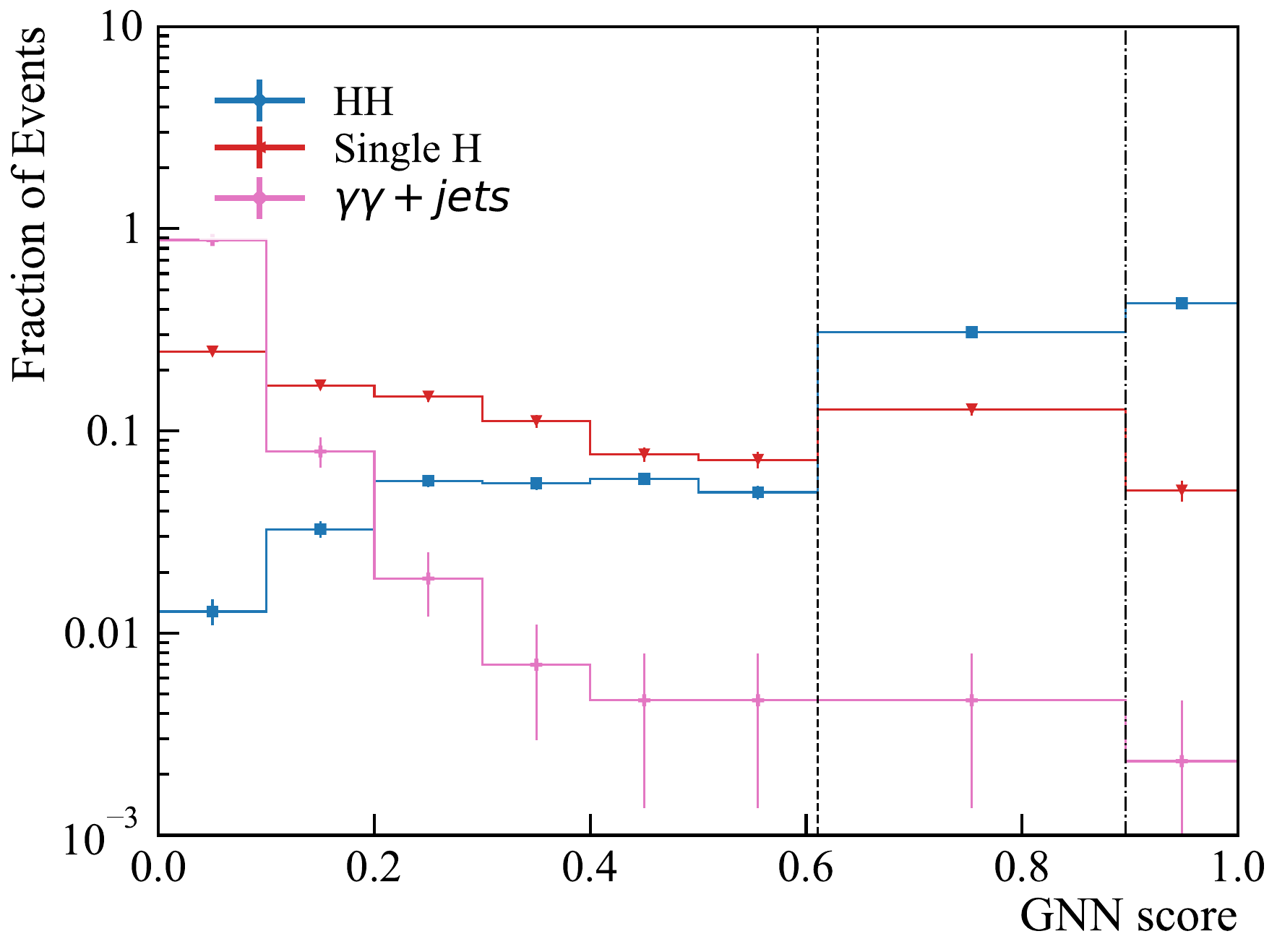}}
    \subfloat[XGBoost Classifier][XGBoost Classifier]{\includegraphics[width=0.5\linewidth]{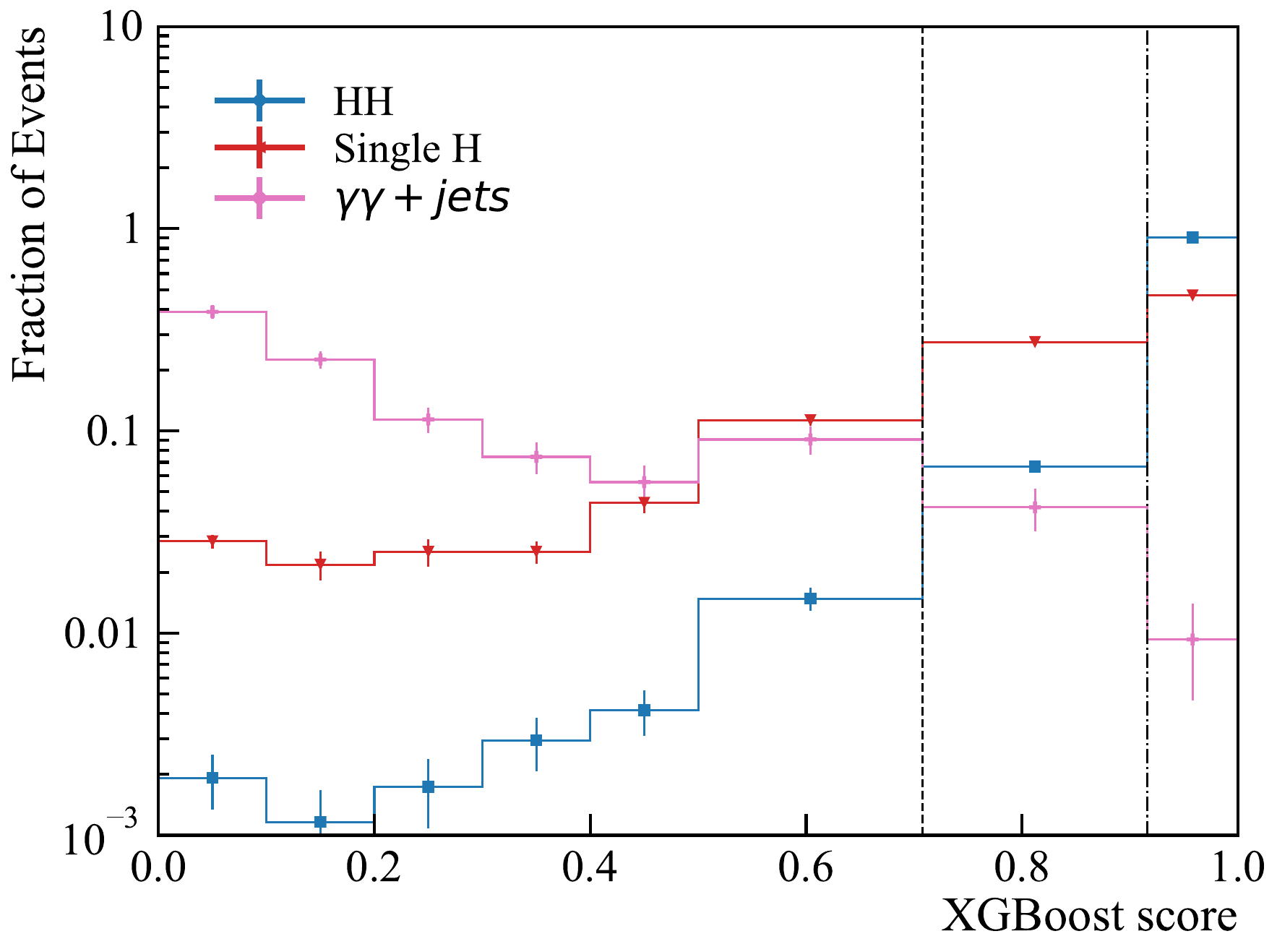}}
    \caption{The classifier output score for both (a) GNN classifier and (b) XGBoost. The dashed vertical lines indicates the optimized binning threshold. In the fit, bins bellow 0.5 are merged in a single bin.}
    \label{fig:score}
\end{figure}

The goal of this analysis is to improve the sensitivity of the non-resonant double Higgs boson production by employing advanced machine learning techniques for signal-vs-background classification. For each trained model, a statistical interpretation is performed based on the classifier output score. As an initial performance comparison, the trained classifiers are evaluated using the Receiver Operating Characteristic (ROC) curve, which quantifies the trade-off between signal efficiency (true positive rate) and background rejection (1 – false positive rate) as a function of the threshold applied to the model output score. Figure~\ref{fig:ROC} demonstrates a significant improvement in both signal efficiency and background rejection when using the GNN model compared to the traditional tree-based classifier. The GNN overperforms the XGBoost  across the full range of classifier thresholds, as evidenced by the ROC curve. An improvement of 18.7\% is observed in the area under the ROC curve (AUC). The observed improvement from the GNN model is attributed to its ability to leverage geometric information and event topology through the graph-based representation. By explicitly encoding spatial relationships and hierarchical connections (decays, interactions, ...) among reconstructed objects, the GNN architecture captures complex inter-object correlations that are not accessible to standard decision tree models.
In addition, we define two bins in the classifier output score above the threshold of 0.5, optimized to maximize the expected statistical significance and signal purity. These bins correspond to score regions where the classifier exhibits the greatest discriminating power. The region below 0.5 is grouped into a single inclusive bin in the fit, as it contributes negligibly to the overall analysis sensitivity. The expected significance is computed using the Asimov approximation to the profile likelihood ratio~\cite{Cowan:2010js}. Each bin is required to contain at least one expected background event. Figure \ref{fig:score} shows the output score distributions for both the GNN and XGBoost classifiers. The vertical dashed lines indicate the boundaries of the two score bins optimized for the statistical analysis. In addition to the ROC comparison, the score distributions further highlight the better separation power of the geometry-aware GNN model relative to the tree-based XGBoost for both single Higgs and the continuum background. 
However, it is important to emphasize that the analysis is limited by the low statistics of the continuum $\gamma\gamma + $ jets background sample. In our simulation, each raw event entry from this background corresponds to approximately 3.23 weighted events after applying cross sections and scaling to the integrated luminosity. Therefore, the requirement of one background event per bin can be considered aggressive, particularly for bins enriched with signal. Despite this limitation, we report results both with and without the inclusion of the continuum $\gamma\gamma + $ jets contribution in the final statistical fit, although it is included in the training of the machine learning models to preserve the dominant background modeling. 

\begin{figure}[ht]
    \centering
    \includegraphics[width=0.85\linewidth]{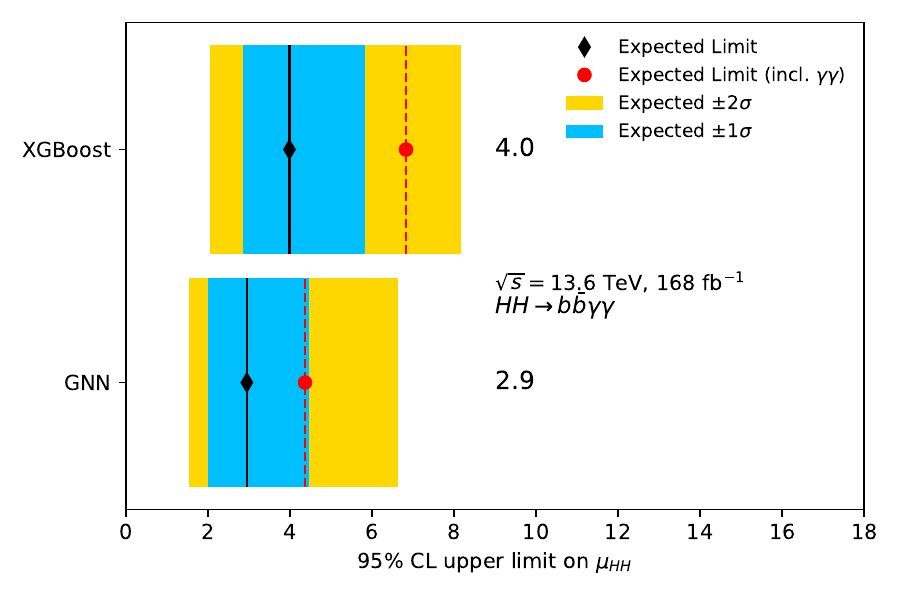}
    \caption{The expected 95\% CL upper limits on $\mu_{HH}$ are presented for both the XGBoost and GNN classifiers with their corresponding $\pm1\sigma$ and $\pm2\sigma$ uncertainty bands. Results obtained when including the $\gamma\gamma +$ jets continuum background are highlighted using red solid markers.}
    \label{fig:limit}
\end{figure}

To quantify the sensitivity of the analysis, we evaluate both the expected discovery significance and the 95\% confidence level (CL) upper exclusion limit \cite{CL}. These statistical metrics are computed using a binned likelihood fit to each classifier output score, implemented via the \texttt{Pyhf} framework~\cite{Heinrich:2021gyp}. A total background normalization uncertainty of 10\% is used as a log-normal nuisance parameter in the fit. The observed number of events is assumed to be equal to the sum of the expected background and the SM HH signal (ggF+VBF). \\
Table \ref{tab:sig} summarizes the expected discovery significance obtained with the fit for both the XGBoost and GNN classifiers, and compares the results to the latest ATLAS values reported in~\cite{atlas_results}.
\begin{table}[hbt!]
    \centering
    \begin{tabular}{lcc}
    \hline \hline
         &  XGBoost  & GNN \\
    \hline      
    Expected significance  & 0.56 & 0.88 \\ 
    
    \hline \hline 
    \end{tabular}
    \caption{Expected discovery significance for both XGBooost and GNN assuming a 10\% background uncertainty.}
    \label{tab:sig}
\end{table}
The expected 95\% CL upper limits on the signal strength, defined as $\mu_{HH} = \sigma(pp \to HH) / \sigma_{\text{SM}}(pp \to HH)$, are shown in Figure~\ref{fig:limit} for each machine learning classifier considered. Limits are presented for two scenarios: one including the continuum $\gamma\gamma + $ jets background (red) and one excluding it (black) to illustrate the impact of limited background statistics. We must stress that even removing the dominant background from the fit, the highest output score bins contain at least 1 total event from the resonant background. The XGBoost classifier yields an expected 95\% CL upper limit of approximately 4.0 times the SM prediction for double Higgs boson production.
In contrast, the graph-based model achieves a significantly tighter constraint, with an expected upper limit of approximately 2.9 times the SM cross-section representing an improvement of nearly 28\% relative to the XGBoost baseline. This gain demonstrates the advantage of representing events as graphs, where the geometric and topological structure of the event is explicitly encoded. By learning from both the features of individual physics objects and their spatial and hierarchical relationships, the GNN is able to capture correlations and event-level dynamics that are not explicitly provided as input features in traditional tree-based classifiers.\\

Our results are comparable to the latest exclusion limits reported by the ATLAS Collaboration, which combine the full Run-2 and partial Run-3 datasets. The ATLAS analysis employs an XGBoost-based classifier together with sophisticated event categorization and a full detector simulation, reporting observed (expected) upper limits on the Higgs boson pair production cross-section of 3.8 (3.7) times the SM prediction~\cite{atlas_results}. Using only Run-3 data, ATLAS sets observed (expected) limits of 5.8 (5.0) times the SM prediction, highlighting the sensitivity improvement achieved with the combined Run-2 and Run-3 dataset. Despite the simplified setup used in this study, the XGBoost-based analysis presented here yields an expected limit that is comparable to the ATLAS Run-3-only result, indicating good consistency with current experimental results. Notably, GNN analysis significantly outperforms the ATLAS Run-3-only expected limit and yields results that are competitive with, and slightly better than, the combined Run-2 + partial Run-3 ATLAS limit at an integrated luminosity of 308~fb\(^{-1}\). By extrapolating our GNN-based limit to 308~fb\(^{-1}\), we obtain an expected upper limit of approximately 2.1 times the SM cross-section, corresponding to a 28\% improvement relative to 168~fb\(^{-1}\) results. While our study is based on simulation and simplified detector modeling and therefore cannot be directly compared to the full experimental results, the relative performance gains between classifier architectures provide valuable insights into the potential of advanced machine learning techniques -- particularly graph-based models -- for future double Higgs boson searches at the LHC and beyond. 

\begin{figure}[hbt!]
    \centering
    \includegraphics[width=0.7\linewidth]{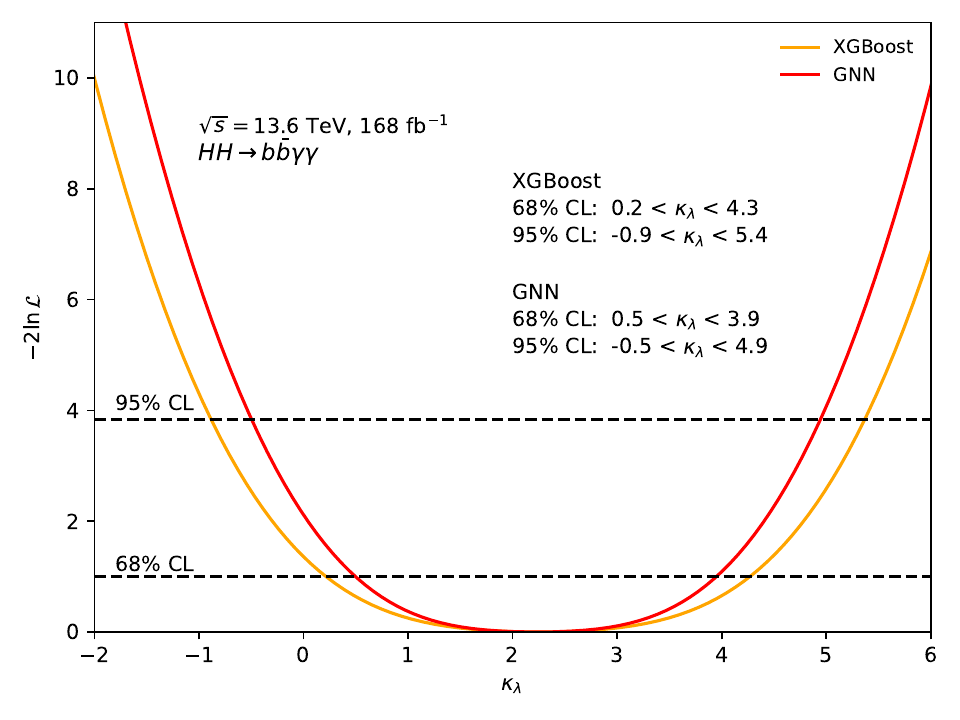}
    \caption{Negative log-likelihood as a function of $\kappa_{\lambda}$ for XGBoost (orange) and GNN (red). Dashes lines represent the 95\% and 68\% limits. }
    \label{fig:kl_scan}
\end{figure}

Although this analysis is not specifically optimized for a measurement of the Higgs boson self-coupling modifier $\kappa_{\lambda}$, we perform a one-dimensional profile likelihood scan to evaluate the 68\% and 95\% CL on $\kappa_{\lambda}$. In this procedure, only $\kappa_{\lambda}$ is treated as a free parameter, while all other Higgs couplings are fixed to their SM values. Importantly, the scan accounts only for the effect of $\kappa_{\lambda}$ on the total production cross-section, without considering any modifications to the kinematic distributions or branching ratios. Figure~\ref{fig:kl_scan} shows the resulting negative log-likelihood profiles as a function of $\kappa_{\lambda}$ for both the XGBoost (orange) and GNN (red) classifiers.
Despite neither model being explicitly trained on BSM signal samples with varying $\kappa_{\lambda}$, both demonstrate sensitivity to deviations from the SM expectation. Notably, the GNN-based analysis yields a significantly tighter confidence interval. The measured 95\% (68\%) confidence interval on the $\kappa_{\lambda}$ modifier, as reported by the ATLAS Collaboration using the full Run-2 and a partial Run-3 dataset, is [-1.7,\ 6.6] ([-0.4,\ 5.1])~\cite{atlas_results}.

\begin{table}[hbt!]
    \centering
    \begin{tabular}{lccc}
    \hline\hline
         & XGBoost & GNN \\
    \hline     
   $\kappa_{\lambda}$ 68\% CL & [0.2, 4.3] & [0.5, 4.0]  \\
   $\kappa_{\lambda}$ 95\% CL & [-0.9, 5.4] & [-0.5, 5.0] \\
    \hline\hline
    \end{tabular}
    \caption{Comparison of the 68\% and 95\% confidence intervals \( \kappa_{\lambda} \) modifier, as obtained from the profile likelihood scan in this analysis using the XGBoost and GNN classifiers.}
    \label{tab:kl_CL}
\end{table}

\begin{figure}[htpb]
    \centering
    \includegraphics[width=0.7\linewidth]{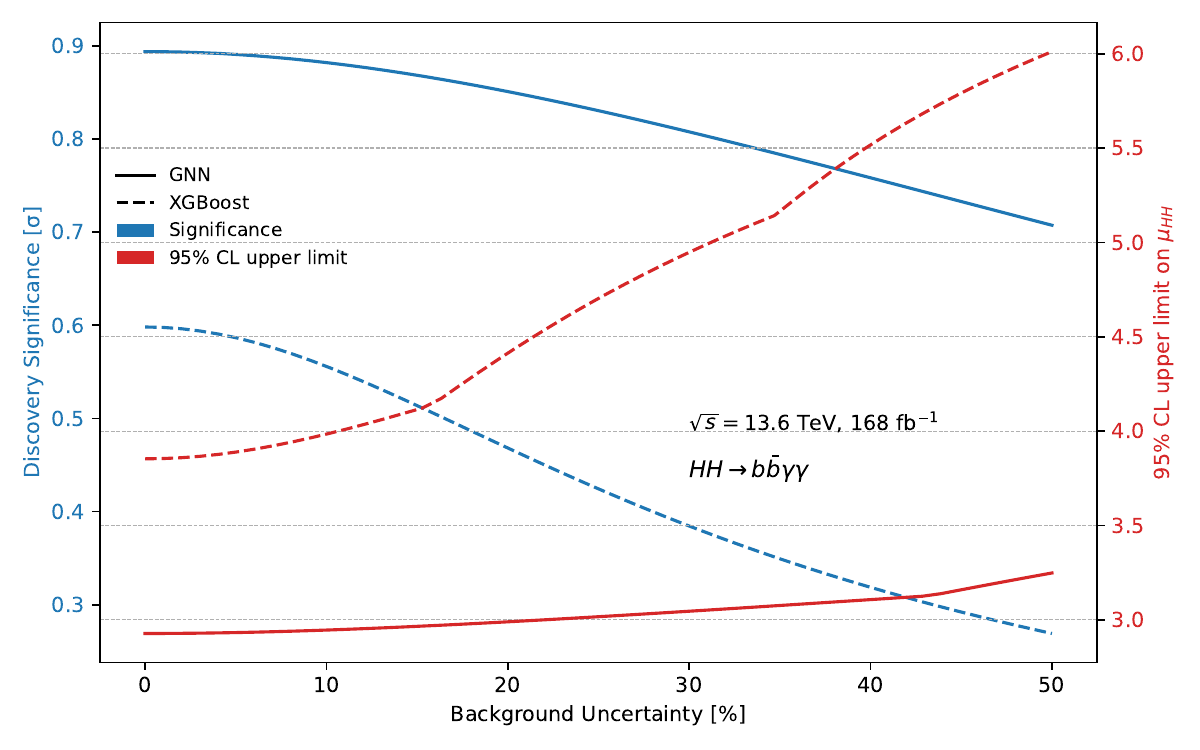}
    \caption{Impact of background uncertainty on the expected discovery significance (blue) and 95\% CL upper limit (red) for the XGBoost (dashed lines) and GNN (solid lines) classifiers.}
    \label{fig:sig_lim_vs_bkgsys}
\end{figure}

To assess the impact of the assumed 10\% background uncertainty on the final results, we perform the statistical analysis under varying background uncertainty scenarios. Figure~\ref{fig:sig_lim_vs_bkgsys} shows the dependence of both the expected discovery significance (in blue) and the expected 95\% CL upper limit  (in red) as a function of the background uncertainty. 
Results are shown for both the XGBoost (solid lines) and GNN (dashed lines) classifiers. This comparison illustrates the robustness of the GNN performance under increased background systematic uncertainty. While both classifiers show a degradation in sensitivity as the background uncertainty increases, the GNN outperforms the XGBoost model across the entire tested range. Thus graph-based models are better able to extract discriminative information from the event topology, making it less sensitive to fluctuations to background normalization.
\section{Conclusion}
\label{sec:conc}
In this work, we explored the potential of advanced machine learning algorithms to enhance the sensitivity of double Higgs boson searches in the $HH \to b\bar{b}\gamma\gamma$ decay channel at the LHC. Despite its low branching ratio, this final state offers a favorable balance of clean photon signatures and high-resolution invariant mass reconstruction, making it a compelling probe of the Higgs boson self-coupling. We implemented and compared two different machine learning algorithms: a tree-based classifier using XGBoost and a GNN that uses geometric event information. The GNN model outperformed the traditional XGBoost classifier, giving an improvement of 28\% in expected upper limits on the signal strength and demonstrating stronger resilience to background systematic uncertainties. Our statistical analysis showed that the GNN classifier achieves an expected 95\% CL upper limit on the di-Higgs cross-section of approximately 2.9 times the SM prediction, compared to 4.0 with XGBoost.

An improvement of nearly 60\% in discovery significance is achieved using the GNN classifier. Furthermore, the GNN enhances sensitivity to the trilinear Higgs self-coupling parameter $\kappa_{\lambda}$, yielding tighter confidence intervals—even without being explicitly trained on BSM signal samples. When compared to the latest ATLAS results, which combine Run-2 and partial Run-3 datasets, the GNN approach shows significant gains in both the expected upper limit on the double Higgs cross-section and the allowed range for $\kappa_{\lambda}$.  This demonstrates the capacity of geometry-based models to generalize and capture features that are indicative of new physics. Overall, results of this paper demonstrate that using geometric learning into collider analyses can significantly enhance the search for rare processes such as the double Higgs production.

\section*{Acknowledgment}
This work is supported by the United Arab Emirates University (UAEU) under UPAR Grant No. 12S162 and Start-Up Grant No 12S157. The authors highly thanks the AI and Robotics Lab of United Arab Emirates University for offering computing facilities including HPC and DGX1 for MC simulation and ML training.

\section*{Data and Code Availability}
The datasets and code used in this analysis can be provided by the corresponding author upon reasonable request.\\

\vspace{0.2cm}
\noindent

\let\doi\relax

\bibliographystyle{utphys}
\bibliography{bib.bib}

\end{document}